\documentclass{iau}
\usepackage{graphicx,natbib}

\title[Moving Mesh Simulations of G2] 
{3D Moving Mesh Simulations of Galactic Center Cloud G2}

\author[P. Chris Fragile et al.]   
{P. Chris Fragile$^1$, Peter Anninos$^2$,
 \and Stephen D. Murray$^2$}

\affiliation{$^1$Department of Physics \& Astronomy, College of Charleston \\ 66 George Street, Charleston, SC 29424, USA \\ email: {\tt fragilep@cofc.edu} \\[\affilskip]
$^2$Lawrence Livermore National Laboratory \\ Livermore, CA 94550, USA}

\pubyear{2013}
\volume{IAU303-GC2013}  
\jname{The Galactic Center: Feeding and Feedback in a Normal Galactic Nucleus}
\begin{document}

\maketitle

\begin{abstract}
Using three-dimensional, moving-mesh simulations, we investigate the future evolution of the recently discovered gas cloud G2 traveling through the galactic center. ÊFrom our simulations we expect an average feeding rate onto Sgr A* in the range of $(5-19) \times 10^{-8} M_\odot\mathrm{~yr}^{-1}$ beginning in 2014. This accretion varies by less than a factor of three on timescales $\sim 1$ month, and shows no more than a factor of 10 difference between the maximum and minimum observed rates within any given model. These rates are comparable to the current estimated accretion rate in the immediate vicinity of Sgr A*, although they represent only a small ($<10$\%) increase over the current expected feeding rate at the effective inner boundary of our simulations ($r_\mathrm{acc} = 750 R_S \sim 10^{15} \mathrm{cm}$). ÊWe also explore multiple possible equations of state to describe the gas.  In examining the Br-$\gamma$ light curves produced from our simulations, we find that all of our isothermal models predict significant (factor of 10) enhancements in the luminosity of G2 as it approaches pericenter, in conflict with observations. ÊModels that instead allow the cloud to heat as it is compressed do better at matching observations.
\end{abstract}

\firstsection 

\section{Simulations}

Using the moving mesh functionality introduced in \cite{anninos12}, we performed a small set of numerical simulations studying the future evolution of a gas cloud on the orbit of G2 through the galactic center.  Each simulation uses a three-dimensional Cartesian grid with a starting size of 14 cloud radii on a side, resolved with 256 zones in each dimension, giving an initial linear resolution of $\Delta x$, $y$, $z = 8.2 \times 10^{13}$ cm.  We are able to study this problem with this relatively small grid because the mesh follows the cloud in its orbit; for simplicity, we fix the motion of the mesh using the Keplerian velocity of the cloud.  We experiment with three different equations of state: isothermal (models with ``i1'' in the name), isentropic (``i53''), and polytropic (``p53''). We also consider two different static models for the background gas (``b1'' and ``b2''). 
Three of the models we discuss were originally presented in \cite{anninos12}, which also goes into much greater depth about the details of the simulations; the one new simulation presented here, cc\_i1\_b2\_95p, uses the orbital parameters of \cite{phifer13}, while the earlier simulations used the parameters of \cite{gillessen12}.


\section{Results}

\begin{figure}[b]
\begin{center}
 \includegraphics[width=2.6in]{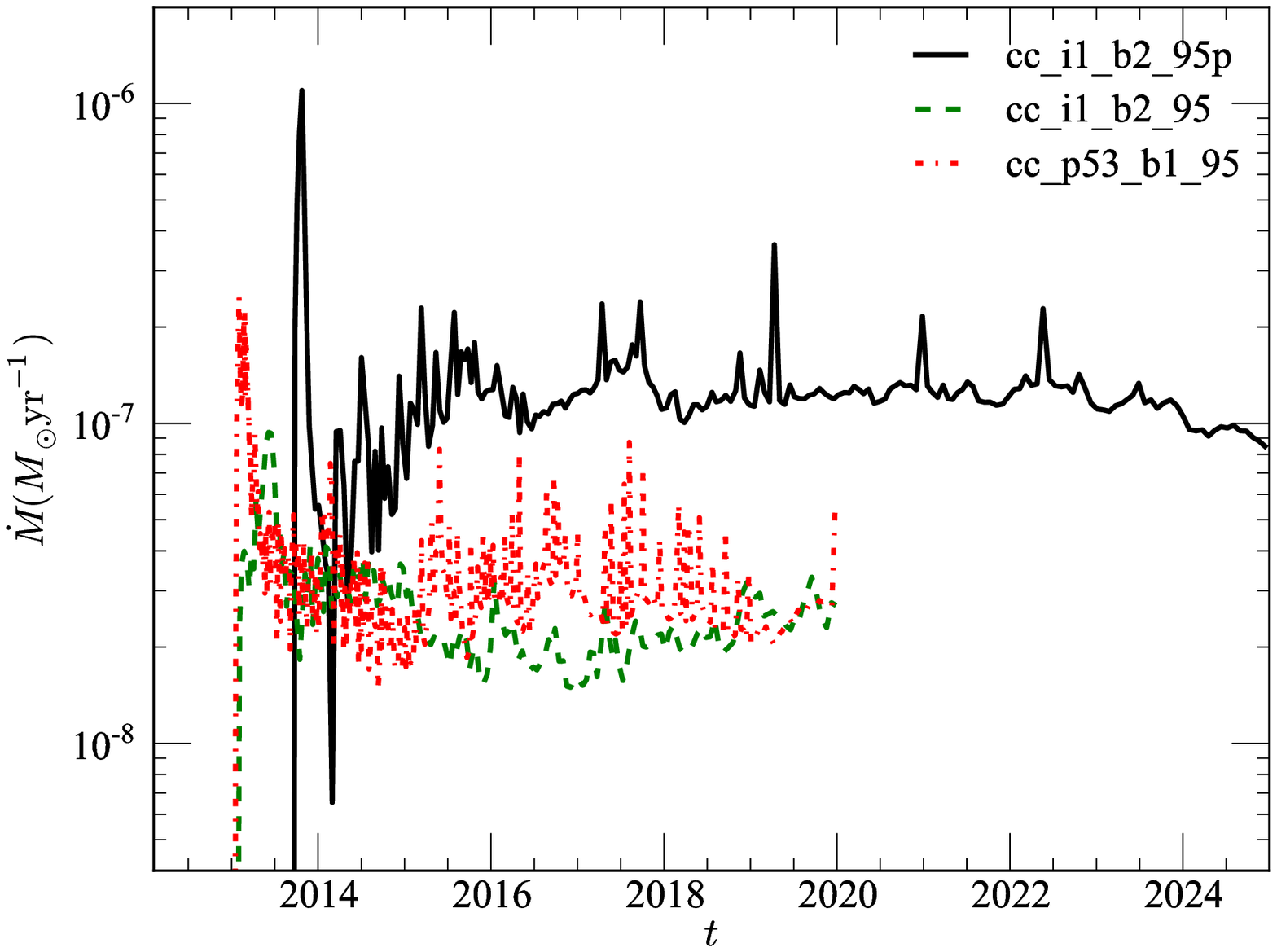}
 \includegraphics[width=2.6in]{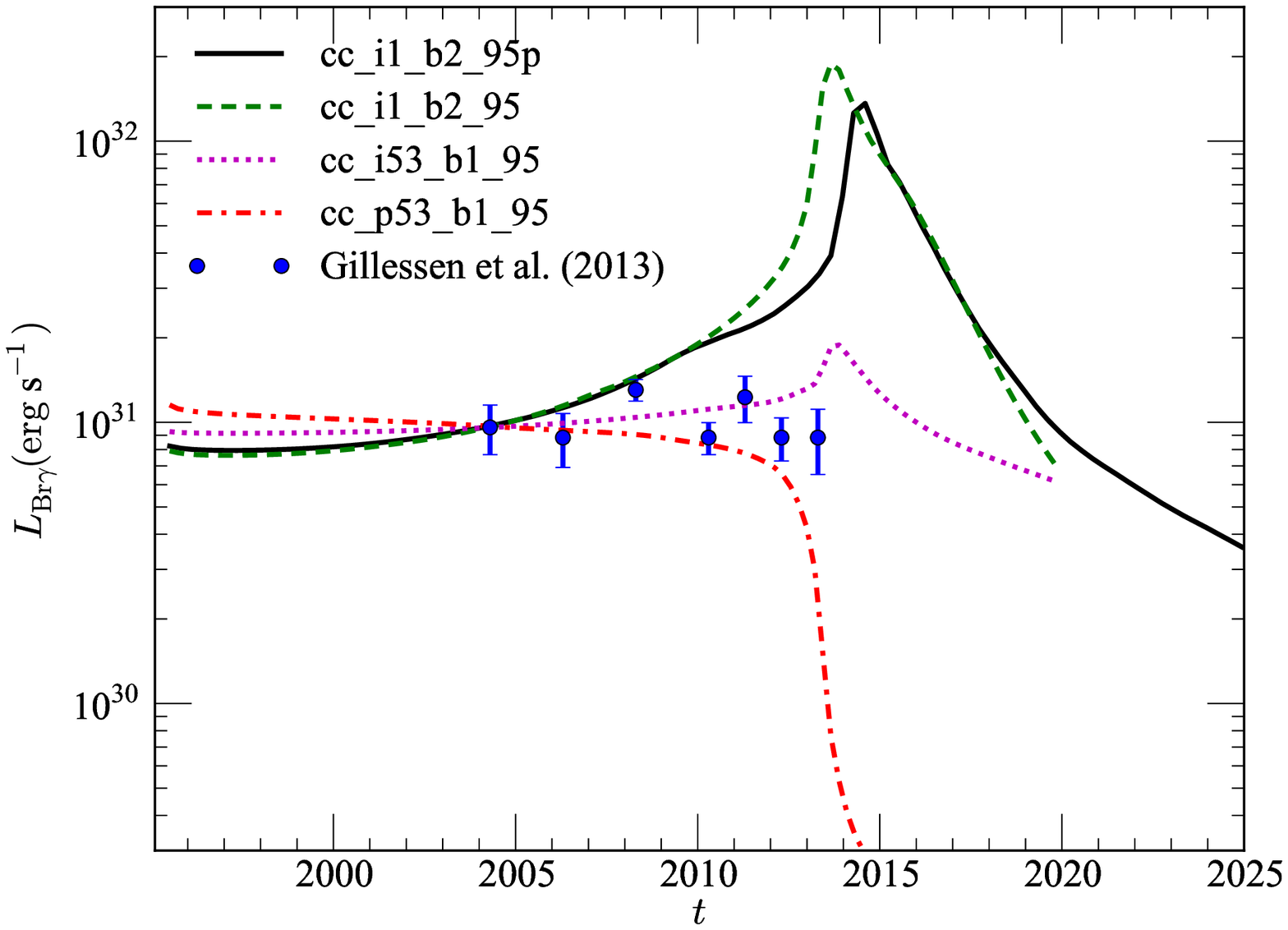} 
 \caption{{\it Left:} Plot of the instantaneous mass accretion rate through $r_\mathrm{acc}$.  {\it Right:} Br-$\gamma$ light curves covering the full time span of our simulations.}
   \label{fig1}
\end{center}
\end{figure}

Figure \ref{fig1} shows the mass accretion rate through the effective inner radius of our simulations, $r_\mathrm{acc}$.  The new simulation, cc\_i1\_b2\_95p, has an accretion rate that is more than twice as large.  This is not surprising given that the orbital parameters of the new simulation include a higher eccentricity and a smaller pericenter distance.  The closer approach naturally promotes greater tidal disruption and enhanced mass accretion.  The near constancy of the mass accretion rates (following their initial spike around 2014.1) is consistent with our previous results and now is extended for an additional five years.

For each of our simulations, we calculate Br-$\gamma$ light curves using the case-B recombination emissivity from \cite{ballone13}.
The light curves for 4 of our models
are shown in the right-hand panel of Figure \ref{fig1}.  The important conclusions are:
\begin{itemize}
\item All isothermal models show dramatic brightening starting around 2005 and continuing into 2014.
\item All models show fading of G2 to below the discovery limits sometime during the next decade. 
\item Only models that allow for substantial heating of the cloud, cc\_i53\_b1\_95 and \\ cc\_p53\_b1\_95, appear to match the data reasonably well past 2008.
\end{itemize}
One important factor that is not considered in this study is the role of magnetic fields.  Strong magnetic fields might allow G2 to resist tidal compression and match observations better \citep{shcherbakov13}.

\begin{acknowledgement}
This work was supported in part by the National Science Foundation under Grant No. NSF PHY-1125915, AST-1211230, and by NSF Cooperative Agreement No. EPS-0919440 that included computing time on the Clemson University Palmetto Cluster.  The work by PA and SDM was performed under the auspices of the U.S. Department of Energy by Lawrence Livermore National Laboratory under Contract DE-AC52-AC52-07NA27344.
\end{acknowledgement}

\end{document}